\documentclass[conference]{IEEEtran}
\IEEEoverridecommandlockouts
\usepackage{cite}
\usepackage{amsmath,amssymb,amsfonts}
\usepackage{graphicx}
\usepackage{textcomp}
\usepackage{xcolor}

\usepackage{algpseudocode}
\usepackage{algorithm}
\usepackage{times}
\usepackage{epsfig}
\usepackage{graphicx}
\usepackage{hyperref}
\usepackage{adjustbox}
\usepackage{multirow}
\usepackage{bbm}
\usepackage{bm}
\usepackage{tablefootnote}
\usepackage{enumitem}
\DeclareMathOperator*{\argmax}{arg\,max}

\def\BibTeX{{\rm B\kern-.05em{\sc i\kern-.025em b}\kern-.08em
    T\kern-.1667em\lower.7ex\hbox{E}\kern-.125emX}}
\begin{document}

\title{Illumination Variation Correction Using Image Synthesis For Unsupervised Domain Adaptive Person Re-Identification
\thanks{This material is based on research sponsored by U.S. Department of Homeland Security (DHS) under Grant Award Number 17STCIN00001. The U.S. Government is authorized to reproduce and distribute reprints for Governmental purposes notwithstanding any copyright notation thereon. The views and conclusions contained herein are those of the authors and should not be interpreted as necessarily representing the official policies or endorsements, either expressed or implied, of DHS or the U.S. Government.
}
}

\author{
\IEEEauthorblockN{Jiaqi Guo}
\IEEEauthorblockA{\textit{School of Electrical and Computer} \\
\textit{Engineering, Purdue University}\\
West Lafayette, IN, USA \\
guo498@purdue.edu}
\and
 \IEEEauthorblockN{Amy R. Reibman}
\IEEEauthorblockA{\textit{School of Electrical and Computer} \\
\textit{Engineering, Purdue University}\\
West Lafayette, IN, USA\\
reibman@purdue.edu}
 \and
\IEEEauthorblockN{Edward J. Delp}
\IEEEauthorblockA{\textit{School of Electrical and Computer} \\
\textit{Engineering, Purdue University}\\
West Lafayette, IN, USA\\
ace@ecn.purdue.edu}
}

\maketitle

\begin{abstract}
Unsupervised domain adaptive (UDA) person re-identification (re-ID) aims to learn identity information from labeled images in source domains and apply it to unlabeled images in a target domain.
One major issue with many unsupervised re-identification methods is that they do not perform well relative to large domain variations such as illumination, viewpoint, and occlusions. 
In this paper, we propose a Synthesis Model Bank (SMB) to deal with illumination variation in unsupervised person re-ID.
The proposed SMB consists of several convolutional neural networks (CNN) for feature extraction and Mahalanobis matrices for distance metrics.
They are trained using synthetic data with different illumination conditions such that their synergistic effect makes the SMB robust against illumination variation.
To better quantify the illumination intensity and improve the quality of synthetic images, we introduce a new 3D virtual-human dataset as source domains for image synthesis.
Both GAN-based and Diffusion-based methods are explored for unpaired image-to-image translation.
From our experiments, the proposed SMB outperforms other synthesis methods on several re-ID benchmarks. 
\end{abstract}


\section{Introduction}
Given a person-of-interest (query image), the goal of person re-identification (re-ID) is to retrieve the images
from a dataset of the same person-of-interest across different cameras at different times and locations (gallery images). 
The person re-ID problem has wide applications in surveillance and security \cite{ye_2021}.
It is also challenging due to viewpoint and illumination variations across cameras, as well as potential occlusions, detection errors, and background clutter in query and gallery images \cite{karanam_2018}. 
With the recent advancement in deep neural networks (DNN), the person re-ID community has achieved very good progress relative to several re-ID benchmarks \cite{sun_2018, zhang_2020, wang_2018, wu_2019}.
In this paper, ``domain'' describes a set of images with similar style (e.g., same illumination, viewpoint, cameras). We will also use the term ``dataset'' to indicate images from a domain.
Most of the above studies were carried out under the supervised settings, where data from the same domain are split into training and test sets. 
However, when the trained DNN models are tested on new domains, there is usually a drastic performance degradation. 
Re-ID data collection and annotation is expensive and labor intensive, which makes supervised methods less desirable.
Therefore, unsupervised domain adaptive (UDA) person re-ID has recently attracted attention \cite{bai_2021,zhong_2019,wei_2018,deng_2018,song_2020}.
In the unsupervised scenario, we have access to one or more labeled domains (the ``source domains''), and unlabeled images from another domain that we wish to use. 
We call this new domain the ``target domain''. 
The goal is to improve the re-ID performance on the target domain.

One approach to unsupervised person re-ID problems is synthesis augmentation \cite{zhong_2019,wei_2018,deng_2018, bak_2018}. 
The synthesis augmentation methods use generative models such as generative adversarial networks (GAN) to create labeled synthetic images from images in the source domains.
The synthetic images approximate the style of the target domain, and are used to fine tune the re-ID models pre-trained on the source domains.
Most of the synthesis augmentation methods directly used the entire source domain in learning the image translation functions \cite{zhong_2019,wei_2018,deng_2018}. 
A few methods take some extra steps of pre-processing the source domain such as subset selection according to illumination \cite{bak_2018}.
In these works, there is a large distribution gap between the source and the target domains when training the generative models.
This makes it difficult to learn the image translation function.

\begin{figure*}[t]
\begin{center}
\includegraphics[width=2\columnwidth]{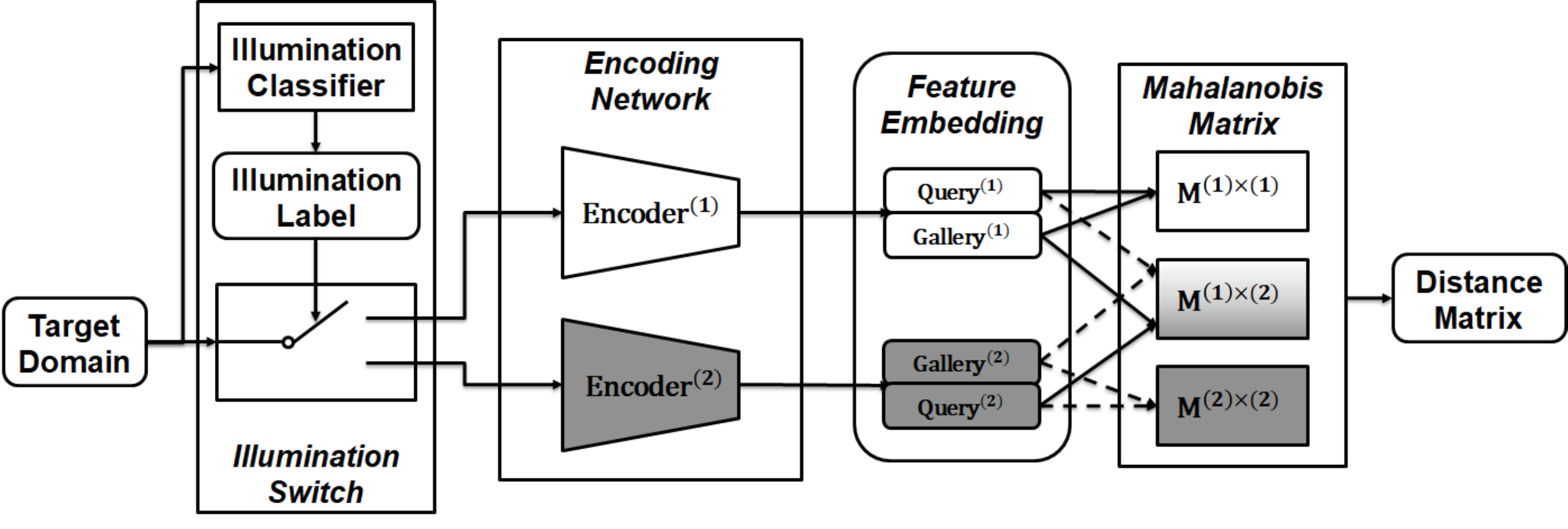}
\caption{
Block diagram of the proposed \textbf{Synthesis Model Bank (SMB)} where we assume there are two illumination conditions in the target domain ($N=2$). 
}
\end{center}
\label{fig_flowchart}
\end{figure*}

In this paper, we use the illumination conditions of the target domain to refine the source domain before training the generative models.
We also propose a deep learning architecture called Synthesis Model Bank (SMB) to deal with illumination variation for unsupervised person re-ID. 
The basic idea of SMB is to use different re-ID models for different illumination conditions.
Figure \ref{fig_flowchart} shows the block diagram of our proposed architecture.
The Synthesis Model Bank consists of three components (we assume that there are $N$ different illumination conditions in the target domain): an Illumination Switch to estimate the illumination labels of input images, $N$ parallel encoding networks that convert the input images into feature vectors, $\frac{N\times (N-1)}{2}+N$ Mahalanobis Matrices that compute the distances between the feature vectors.
Each encoding network is trained for one of the $N$ most common illumination conditions in the target domain.
Based on the estimated illumination labels, we select the corresponding encoding network and Mahalanobis Matrix to compute the distances between query and gallery images for identity retrieval.
The definition of illumination labels will be discussed in Section \ref{subsubsec_uliri}.

To generate the synthetic images for training the encoding networks and the Mahalanobis Matrices,
we introduce a new virtual-human image dataset as the input to generative models, the \textbf{U}E4 \textbf{L}abeled \textbf{I}mages for Person \textbf{R}e-\textbf{I}dentification (ULI-RI).
The ULI-RI is created using Unreal Engine 4 (UE4) to simulate various outdoor environments with 115 high-quality 3D human models. 
Images in the ULI-RI are labeled according to both illumination intensity and person identity.
Compared with other virtual-human datasets \cite{bak_2018} where the illumination is indirectly controlled by environmental maps, time, or weather, the illumination in our ULI-RI is directly controlled using the \textbf{UE4 SunSky} asset such that the illumination labels is quantitatively linked to the illumination intensity (see Section \ref{subsubsec_uliri} for details).

Our major contributions in this paper can be summarized as follows:
\begin{itemize}
\item We propose a new deep learning architecture called Synthesis Model Bank (SMB) to deal with illumination variation for unsupervised person re-ID.
\item We introduce a new 3D virtual-human dataset for person re-ID which have images labeled by illumination intensity as well as person identity, person rotation angle, and background. 
\item Our experiments show that the proposed SMB outperforms other synthesis methods on several re-ID benchmarks. 
\end{itemize}

\section{Related Work}\label{sec_related}

\begin{table*}[t]
\centering
\caption{3D virtual-human datasets for person re-ID }\label{table_virtual_human}
\begin{adjustbox}{width=1.9\columnwidth}
\begin{tabular}{|c|c|c|c|c|c|c|}
\hline
  & Software & Identity & Background  & Viewpoint  & Illumination & Bounding Box \\
\hline
\hline
 SOMAset & Blender & 50 & \multicolumn{3}{c|}{\centering{Not Controlled$^a$}} &100'000\\
\hline
 PersonX & Unity & 1266 & 6  &36 & Not Controlled & 273'456\\
\hline
 GPR+ & GTA5 & 808 & Not Controlled  &12 & 49$^b$ & 475'104\\
\hline
 SyRI & UE4 & 100 & \multicolumn{2}{c|}{4$^c$} & 140 & 56'000\\
\hline
 ULI-RI & UE4 & 115 & 8  &8 & 8 & 58'880\\
\hline
\end{tabular}
\end{adjustbox}\\
\footnotesize{
$^a$ Each 3D human model (identity) has 8 clothing sets and 250 poses, and is placed in a random environment for rendering.\\
$^b$ The authors selected 7 daytime and 7 weathers to have a total of 49 different illumination conditions. \\
$^c$ Both background and viewpoint of SyRI are controlled by fixing the 3D human models and rotating the virtual cameras.
}
\end{table*}

\subsection{Deep Neural Networks for Person Re-ID}
With the recent advancement in deep neural networks (DNN), the person re-ID performance has achieved very good progress on several benchmark tests \cite{sun_2018, zhang_2020, wang_2018, wu_2019}.
A re-ID system based on deep neural networks typically includes feature extraction learning \cite{ye_2021}, which aims to learn deep feature embeddings/vectors from input images.
Since deep neural networks are initially designed for image classification problems \cite{he_2016,szegedy_2015,bengio_2015,krizhevsky_2012},
many deep re-ID networks follow the classification framework and view the training procedures of person re-ID as a classification problem, where each person (identity) in the dataset is treated as a unique class. 
Usually an extra fully connected layer with Batch Normalization is added before the CrossEntropy layer to define the feature vectors \cite{sun_2018, zhong_2017, sun_2017, ye_2018, han_2023}. 
In this paper, we follow the practice in \cite{bak_2018} and use  ResNet-50 \cite{he_2016} for the encoding networks in our proposed Synthesis Model Bank.

\subsection{Synthesis Augmentation for Unsupervised Person Re-ID}

Since generative adversarial networks (GANs) \cite{goodfellow_2014} were first introduced, using synthetic data for fine-tuning deep neural networks has drawn significant attention. 
An important direction to improve person re-ID performance in unsupervised settings is through synthesis augmentation. 
GANs are used to learn the style of the target domain, and transfer the style to source domains to create labeled synthetic images.  
In \cite{deng_2018}, the authors proposed SPGAN which captures self-similarity and domain dissimilarity to preserve image quality after GAN translation.
In \cite{wei_2018}, Wei et al. explored semantic segmentation map to introduce an extra regularization term in the loss function and proposed PTGAN. 
In \cite{zhong_2019}, Zhong et al. explored a camera-specific strategy to transfer one source domain to every sub-target domain.
In \cite{bak_2018}, Bak et al. proposed a method which first selects the subset of a virtual source domain according to the illumination of the target domain, and then used CycleGAN \cite{zhu_2017} to generate synthetic images from the selected subset.
The method in \cite{bak_2018} assumed there was only one illumination condition in each target domain, whereas images from real-life surveillance cameras often have more complicated illumination variations.
In this paper, we extend this method by including multiple illumination conditions and building a bank of convolutional neural networks and distance metrics to address the illumination variation in the target domain.
We also explore a diffusion based method for unpaired image-to-image translation in comparison to CycleGAN\cite{goodfellow_2014, zhu_2017} used in \cite{bak_2018}.

\subsection{Virtual-Human Datasets for Person Re-ID} \label{subsection_virtual_dataset}

Virtual-human images are typically rendered using a game engine with 3D human and scene models \cite{bak_2018, xiaoxiao_2019},
or the screenshots of human characters from a video game \cite{xiang_2021}. 
In this paper, the virtual-human images are the source domains and used as input of generative models to generate the synthetic images for training the proposed Synthesis Model Bank.

We summarize the existing 3D virtual-human datasets for person re-ID in Table \ref{table_virtual_human}.
SOMAset \cite{igor_2017} contains 50 identities and is designed for re-identifying the same person with different clothing and accessories.
PersonX \cite{xiaoxiao_2019} is specially created to study the effects of viewpoint variation. 
Both SOMAset and PersonX do not include illumination variations when generating the images.
In GPR+ \cite{xiang_2021}, the authors controlled the daytime and the weather in a video game (Grand Theft Auto 5) to obtain various illumination conditions.
Each combination of daytime and weather is assigned a different illumination label. 
In SyRI \cite{bak_2018}, the authors used 140 High Dynamic Range (HDR) environment maps as background and light sources to render the virtual scenes using the UE4 game engine. 
The images from the same HDR map were assigned the same illumination label.
However, this can be inaccurate because the angle between the UE4 virtual cameras and the light sources in the HDR environments causes noticeable illumination variations even for images from the same HDR map.

Note that the illumination conditions in GPR+ and SyRI are indirectly controlled through daytime, weather, or environment maps. 
It is difficult to interpret the brightness or ``illumination intensity'' from the ``illumination labels'' in GPR+ and SyRI. 
In this paper, we introduce a new virtual human dataset called ULI-RI. 
The illumination in the ULI-RI dataset is directly controlled using the \textbf{SunSky} asset of UE4 such that the illumination labels are linked to the illumination intensity through a designed function. Figure \ref{fig_uliri_samples} shows an example of the same identity with different illumination labels in the ULI-RI dataset. 
More detailed description about the ULI-RI dataset and the illumination labels can be found in Section \ref{subsubsec_uliri}.

\section{Our Proposed Approach} \label{sec_method}

In unsupervised domain adaptive person re-ID, we have access to one or several labeled source domains, and aim to achieve better re-ID performance on a real, unlabeled target domain.
The source domain used in this paper mainly refers to the virtual human images generated using the gaming engine, Unreal Engine 4 (UE4) \cite{unreal4}.
The remaining parts of this section will be arranged as follows: we will first describe how the virtual human images are created; next, we will discuss the unpaired image-to-image translation methods to generate the synthetic images; finally, we introduce our proposed Synthesis Model Bank.

\subsection{\textbf{U}E4 \textbf{L}abeled \textbf{I}mages for Person \textbf{R}e-\textbf{I}dentification (ULI-RI Dataset)} \label{subsubsec_uliri}

\begin{figure}[t]
\begin{center}
\includegraphics[width=1\columnwidth]{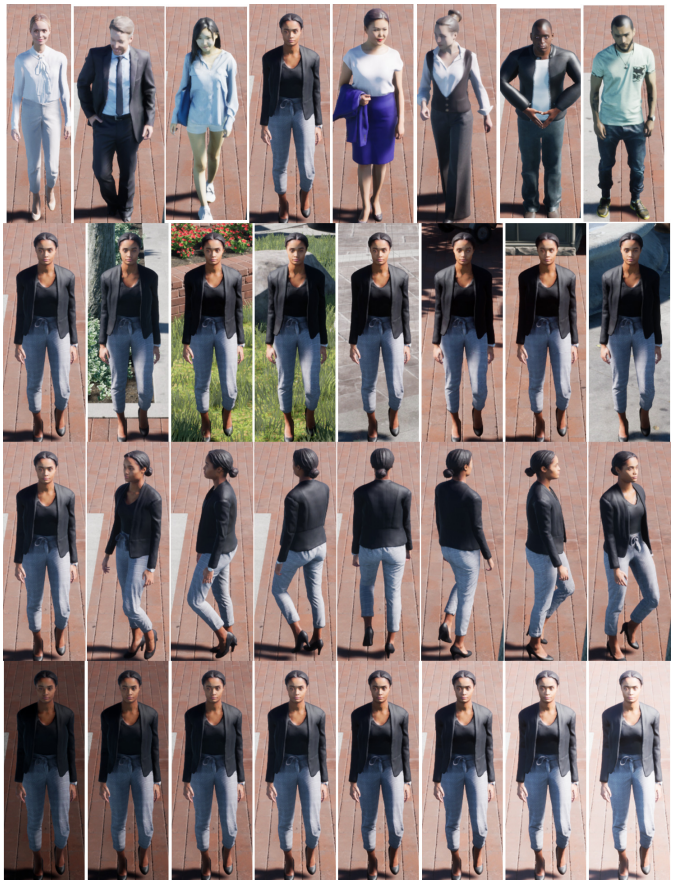}
\end{center}
\caption{Example images from the ULI-RI dataset 
First row: 8 different identities in the ULRI-RI 
Second row: the same identity with 8 different backgrounds (cameras)
Third row: the same identity with 8 different model z-rotation angle (viewpoints)
Fourth row: the same identity under 8 different illumination conditions}
\label{fig_uliri_samples}
\end{figure}

To better quantify the illumination intensity in the target domain and improve the quality of synthetic images, we introduce the \textbf{U}E4 \textbf{L}abeled \textbf{I}mages for Person \textbf{R}e-\textbf{I}dentification (ULI-RI dataset), a new virtual-human image dataset for person re-ID.
The ULI-RI dataset is the virtual source domain (denoted as $\mathbb{V}$) which are used as input to generate the synthetic images by learning the style of the target domain. 
Example images from the ULI-RI are shown in Figure \ref{fig_uliri_samples}. 
The ULI-RI dataset is generated using Unreal Engine 4 to simulate various outdoor environments with 115 high-quality 3D human models. 
The images in the ULI-RI dataset are controlled and labeled by illumination intensity as well as person identity, model z-rotation angle, and background.
The ULI-RI dataset is available at \url{https://lorenz.ecn.purdue.edu/~guo498/ssiai2024}.

\begin{figure}[t]
\begin{center}
\includegraphics[width=1\columnwidth]{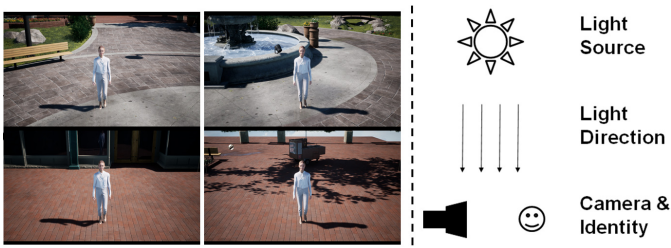}
\end{center}
\caption{Illustration of the camera setup in the virtual environment. Left: example images of the selected locations; Right: relative camera-light source locations }
\label{fig_uliri_cameras_backgrounds}
\end{figure}

\noindent \textbf{Background}: 
We use the \textit{Downtown West Modular Pack} \cite{ue4market} to setup the virtual environment.
This simulates a large outdoor shopping mall and its surrounding areas, with hundreds of high quality models of buildings, gardens, decorations and other construction pieces.
We select 8 locations (examples shown in Figure \ref{fig_uliri_cameras_backgrounds} Left) in the environment to set up  virtual surveillance cameras such that 1) the images have less shadows 2) the camera facing direction is vertical to the light direction (as shown in Figure \ref{fig_uliri_cameras_backgrounds} Right); 3) the backgrounds are different.
The first two rules make sure that under the same UE4 light settings, the person in different images have similar illumination conditions. 
The third rule makes sure that there are diverse backgrounds in the ULI-RI dataset.

\noindent \textbf{Model Z-Rotation Angle (Viewpoint)}:
The viewpoint refers to the angle between the person facing direction and the virtual camera facing direction.
In UE4, this is controlled by the human model z-rotation angle.
For every background and illumination condition, we set 8 values for model z-rotation angle ranging from 0$^\circ$ to 315$^\circ$ in step of 45$^\circ$.
We assign z-rotation label ``0'' to the images where the person is facing the camera, and the rest of the z-rotation labels are assigned in sequence as the person rotates clockwise along its z-axis.
The third row in Figure \ref{fig_uliri_samples} gives an example of the same identity at 8 different model z-rotation angles (background and illumination fixed), and from left to right their z-rotation labels are 0 to 7.

\noindent \textbf{Illumination Label}:
We use the SunSky asset in UE4 to achieve a direct link between the illumination labels and the illumination intensities.
We set the SunSky asset as the only light source in the UE4 virtual environment.
The illumination intensity is adjusted through the \textbf{light intensity parameter} of  ``DirectionalLight'' component of the SunSky asset. 
We use an adjusted exponential function to map the illumination label to the light intensity parameter since it mostly follows the Weber's law \cite{weber_1995} and shows linear changes from human perceptions. 
Mathematically, given an illumination label $I \in \{0,1,...,7\}$, the \textbf{light intensity parameter} of SunSky ($L$) is set by

\begin{equation*}\label{eq_light_label_intensity}
\begin{aligned}
L = exp(I\times 0.5 + 0.6) - 1 \; ;
\end{aligned}
\end{equation*}

\noindent We set 8 levels of illumination intensity in the ULI-RI dataset. 
The last row of Figure \ref{fig_uliri_samples} gives an example of the same identity with 8 different illumination labels (background and z-rotation fixed).
From left to right their illumination labels are 0 to 7.

\noindent \textbf{3D Human Models}:
There are 115 3D human models (identities) in the ULI-RI dataset, which are selected to cover a wide range of gender, race, age, clothing, and accessories. 
Most of the models are collected from online resources such as \href{https://www.unrealengine.com/marketplace/en-US/store}{UE Marketplace}. 
When taking the photos, the 3D human models are placed at the center of the camera view, and we rotate the models along their z-axis to get different viewpoints. 
Every time the illumination intensity, background (camera), or model z-rotation angle is changed, we re-rendered the entire environment to generate more realistic 3D human images. 
In conclusion, we create a total of 115 (identity) $\times$ 8 (background) $\times$ 8 (z-rotation) $\times$ 8 (illumination condition) = 58,880 images, 512 for each identity.

\subsection{Generate Synthetic Images for Training} \label{subsec_2phase}

\begin{figure}[t]
\begin{center}
\includegraphics[width=0.9\columnwidth]{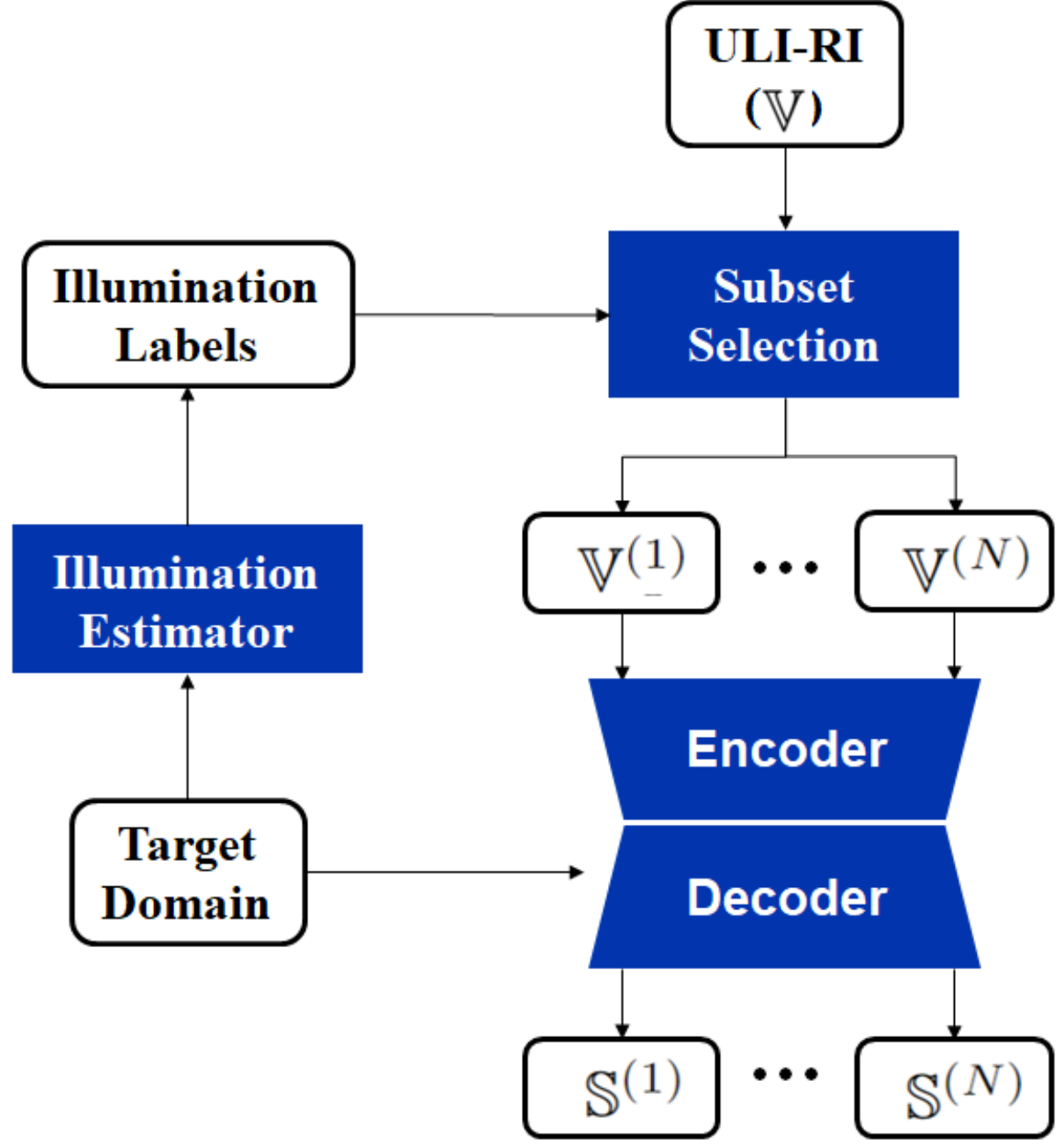}
\end{center}
\caption{ Generating synthetic images using the ULI-RI dataset ($\mathbb{V}$). 
$\mathbb{V}^{(n)}$ is the subset of $\mathbb{V}$ whose illumination label is $n$. $\mathbb{S}^{(n)}$ is the set of synthetic images generated using $\mathbb{V}^{(n)}$. }
\label{fig_synthesis_flowchart}
\end{figure}

To train the proposed Synthesis Model Bank (SMB), we need to generate synthetic images that approximate the style of the target domain.
As illustrated in Figure \ref{fig_synthesis_flowchart}, our approach can be summarized into three steps:
1) Find the $N$ most common illumination labels in the target domain;
2) Select $N$ subsets of ULI-RI according to the most common illumination labels.
Each selected subset will be used as an independent source domain denoted as $\mathbb{V}^{(n)}$ (for $n \in \{1,2,\cdots,N\}$);
3) Use generative models to translate each selected source domain $\mathbb{V}^{(n)}$ into a set of synthetic images (denoted as $\mathbb{S}^{(n)}$).

The \textbf{Illumination Estimator} in Figure \ref{fig_synthesis_flowchart} (denoted as $\mathcal{IE}(\cdot)$) finds the illumination labels for images of the target domain (denoted as $\mathbb{T}$): $$\mathcal{IE}: \mathbb{T}\rightarrow \{1,2,...,8\}$$
$\mathcal{IE}(\cdot)$ is a ResNet-18 classifier \cite{he_2016} which is trained using images and their illumination labels of the ULI-RI dataset.
The most common illumination label $n^{\star}$ in  target domain $\mathbb{T}$ is defined as:
\begin{equation}\label{eq_most_illu}
\begin{aligned}
n^{\star} = \argmax_{n \in \{1,2,...,8 \} } \sum_{t_i \in \mathbb{T}} \mathbbm{1}( \mathcal{IE}(t_i)==n ) \; ,
\end{aligned}
\end{equation}
\noindent where $ \mathbbm{1}(\cdot)$ is the indicator function, $t_i$ is the image in target domain $\mathbb{T}$.
The subset of ULI-RI that has illumination label $n^{\star}$ is selected to create the first source domain, $\mathbb{V}^{(1)}$.
Similarly, we create the source domains $\mathbb{V}^{(2)}, \cdots, \mathbb{V}^{(N)}$ according to the remaining $N-1$ most common illumination labels of the target domain.
From our experiments, most images in a target domain (over $85\%$) are covered by the two most common illumination labels.
Therefore, we set $N=2$ in this paper, i.e., we assume there are two major illumination conditions in a target domain.

\begin{figure}[t]
\begin{center}
\includegraphics[width=1\columnwidth]{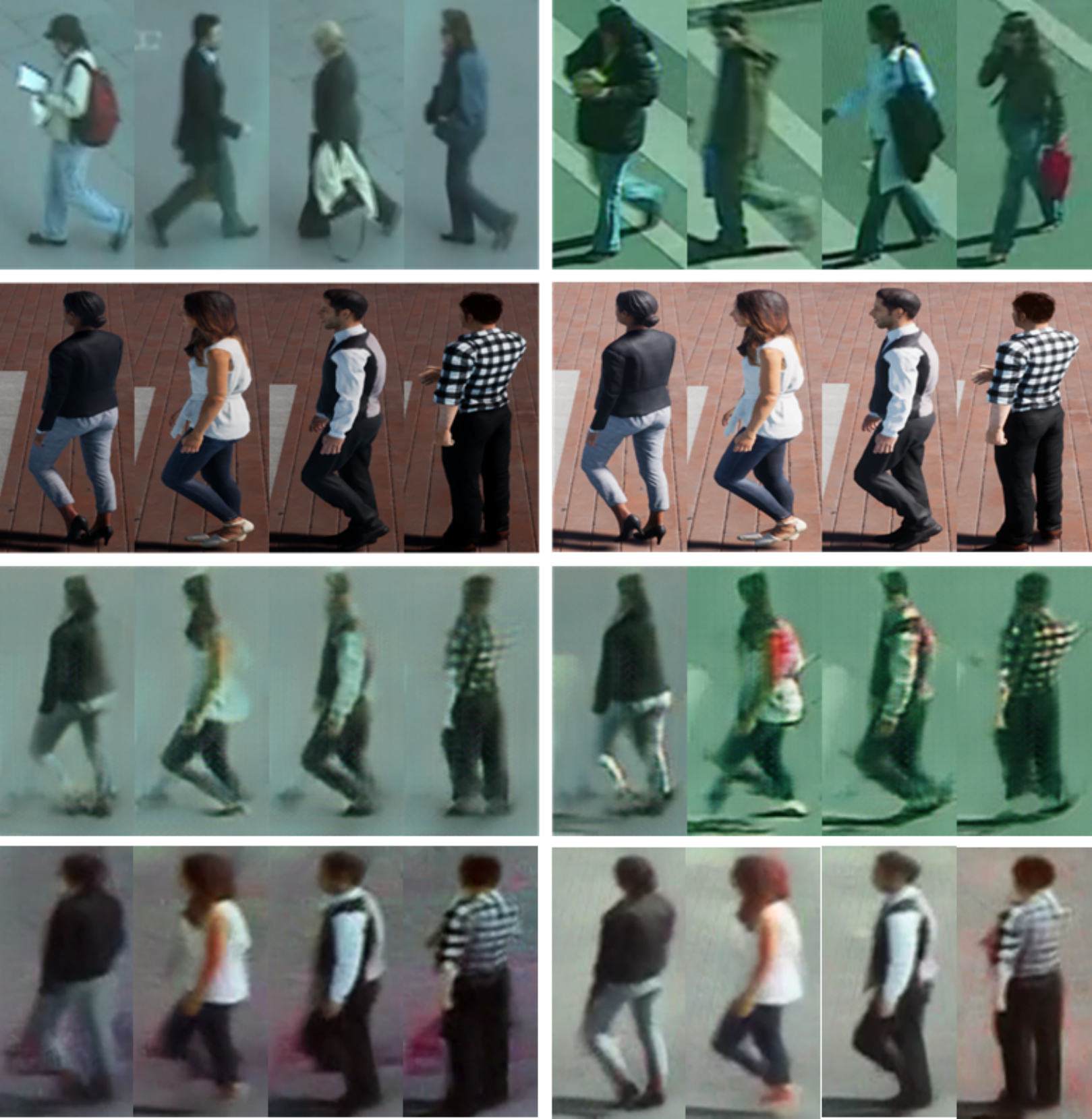}
\end{center}
\caption{First row: images from the target domain, PRID2011 dataset ($t_i\in \mathbb{T}$). Second row: images from the source domains, $\mathbb{V}^{(1)} $ (left) and $\mathbb{V}^{(2)}$ (right). Both are subsets of the ULI-RI dataset.
Third row: synthetic images generated using CycleGAN, $\mathbb{S}^{(1)}_{GAN} $ (left) and $\mathbb{S}^{(2)}_{GAN}$ (right).
Fourth row: synthetic images generated using CycleDiffusion, $\mathbb{S}^{(1)}_{Dif} $ (left) and $\mathbb{S}^{(2)}_{Dif}$ (right).
The target domain images on the left and those on the right have different illumination conditions. 
}\label{fig_prid_synthetic_samples}
\end{figure} 

Given a source domain $\mathbb{V}^{(n)}$ ($n \in \{1,2,\cdots,N\}$) and a target domain $\mathbb{T}$, 
the next step is to learn an image-to-image translation function $G(\cdot)$ that maps $\mathbb{V}^{(n)}$ to $\mathbb{T}$ while preserving important semantic information: $G: \mathbb{V}^{(n)} \rightarrow \mathbb{T}$.
The translated images, i.e. $G(\mathbb{V}^{(n)})$, are denoted as the synthetic images.
Note that the images from $\mathbb{V}^{(n)}$ and $\mathbb{T}$ are not paired. 
There can be infinite number of possible translation functions \cite{xie_2022}.
In this paper, we explore two approaches of learning the translation function, namely CycleGAN \cite{zhu_2017} and CycleDiffusion \cite{wu_2023}.

CycleGAN learns two translation functions $G: \mathbb{V} \rightarrow \mathbb{T}$ and $F: \mathbb{T} \rightarrow \mathbb{V}$ that can transfer the styles between the two domains \cite{zhu_2017}.
In this paper, the translation function $G: \mathbb{V} \rightarrow \mathbb{T}$, which transfers the images in the virtual source domains to the target domain, is needed for generating the synthetic images. 
Two adversarial networks $D_{\mathbb{V}}$ and $D_{\mathbb{T}}$ acting as discriminators are trained along with $G$ and $F$. 
$D_{\mathbb{V}}$ aims to discriminate the images in the virtual source domain from images generated by $F$. 
Analogously, $D_{\mathbb{T}}$ aims to discriminate the images in the target domain from images generated by $G$.

CycleDiffusion \cite{wu_2023} first learns two Denoising Diffusion Probabilistic/Implicit Models (DDPM/DDIM) \cite{ho_2020,song_2022} for source domain and target domain independently. 
We denote the two DDIM as $\mathbf{\mu}_{\mathbb{V}}$ and $\mathbf{\mu}_{\mathbb{T}}$.
Here we use the mean predictor $\mathbf{\mu}$ to denote the neural network instead of the noise predictor $\mathbf{\epsilon}$ as in the original DDPM paper \cite{ho_2020,song_2022}, but they are essentially the same.
Given a noise image $\mathbf{v}_T \sim \mathcal{N}(\mathbf{0}, \mathbf{I})$, $\mathbf{\mu}_{\mathbb{V}}$ generates an image $\mathbf{v}_0$ of source domain $\mathbb{V}$ through the diffusion steps: $$\mathbf{v}_t \sim \mathcal{N}(\mathbf{\mu}_{\mathbb{V}}(\mathbf{v}_{t-1},t), diag(\mathbf{\sigma}_t^2)), \;\; t \in \{T,T-1,\cdots,1\}$$
Similarly, $\mathbf{\mu}_{\mathbb{T}}$ can generate an image of target domain $\mathbb{T}$ using the same noise image $\mathbf{v}_T$.
The CycleDiffusion is developed based on the observation that given two DDPM/DDIM, a fixed latent code can produce ``similar images'' \cite{wu_2023}.
The encoder of CycleDiffusion is defined using $\mathbf{\mu}_{\mathbb{V}}$: 

\begin{equation}\label{eq_latent_code}
\begin{aligned}
\mathbf{z}:= (\mathbf{v_T} \oplus \mathbf{\epsilon_T} \oplus \cdots \oplus \mathbf{\epsilon_1})\\
\mathbf{\epsilon_t} := (\mathbf{v}_{t-1} - \mathbf{\mu}_{\mathbb{V}}(\mathbf{v}_t,t))/\mathbf{\sigma_t}
\end{aligned}
\end{equation}
where $\oplus$ is concatenation along channel dimmension, $T$ is the total number of diffusion steps, $\mathbf{v}_t$ can be computed from the posterior distribution $q(\mathbf{v}_{1:T}| \mathbf{v}_0)$ \cite{song_2022,ho_2020}, $\mathbf{\epsilon}_t$ is the ideal variance to achieve perfect reconstruction (compared with the sampling algorithm in \cite{ho_2020}, where the variance is sampled from a Gaussian distribution), $\mathbf{z}$ is the latent code.
For decoding, we use $\mathbf{v}_T$ as the noise image and $\mathbf{\epsilon}_t$ as the variance in each denoising step of $\mathbf{\mu}_{\mathbb{T}}$. 
The CycleDiffusion sampling algorithm including both encoding and decoding is shown in Algorithm \ref{alg_cyclediffusion}.

\begin{algorithm}
\caption{CycleDiffusion for Unpaired Image-to-Image Translation}
\label{alg_cyclediffusion} 
\begin{algorithmic}[1]
\State \textbf{Input:} Image $\mathbf{v}_0$ from source domain $\mathbb{V}$; DDIM $\mathbf{\mu}_{\mathbb{V}}$ for source domain; DDIM $\mathbf{\mu}_{\mathbb{T}}$ for target domain; encoding steps $T_{es}\leq T$

\State Sample noisy image $\mathbf{t}_{T_{es}} := \mathbf{v}_{T_{es}} \sim q(\mathbf{v}_{T_{es}} \mid \mathbf{v}_{0})$
\For {$t=T_{es},\cdots,1$}
    \State $\mathbf{v}_{t-1} \sim q(\mathbf{v}_{t-1} \mid \mathbf{v}_{t}, \mathbf{v}_{0})$
    \State $\mathbf{\epsilon_t} = (\mathbf{v}_{t-1} - \mathbf{\mu}_{\mathbb{V}}(\mathbf{v}_t,t))/\mathbf{\sigma_t}$
    \State $\mathbf{t}_{t-1}  = \mathbf{\mu}_{\mathbb{T}}(\mathbf{t}_t,t)+\mathbf{\sigma_t}\mathbf{\epsilon_t}$
\EndFor
\State \textbf{Output:} Synthetic image $\mathbf{t_0}$ of the target domain $\mathbb{T}$
\end{algorithmic}
\end{algorithm}

Figure \ref{fig_prid_synthetic_samples} shows the synthetic images generated using CycleGAN and CycleDiffusion respectively. 
Here the target domain is the PRID dataset \cite{hirzer_2011}, which has two major illumination conditions ($N=2$). 
The synthetic images generated using CycleGAN and CycleDiffusion are shown in the last two rows of Figure \ref{fig_prid_synthetic_samples}).
For both generative methods, the generated synthetic images have similar styles and illumination as the target domain, 
and the identity information are kept from the source domain such that they have the same identity labels as their source images.
Compared with CycleGAN, the synthetic images generated by CycleDiffusion have less distortion or missing part. 
We provide a quantitative evaluation on the two generative methods in Section \ref{subsec_gan_diffusion}.

\subsection{Synthesis Model Bank}

The basic concept of the proposed Synthesis Model Bank (SMB) for re-ID is to train multiple re-ID models, and use the most suitable models according to the illumination conditions of the input images.
As shown in Figure \ref{fig_flowchart}, the proposed SMB consists of three components,
1) an Illumination Switch that classifies the query image and the gallery image into one of $N$ illumination conditions,
2) $N$ encoding networks that convert input images into feature vectors for a specific illumination condition.
3) $\frac{N\times (N-1)}{2}+N$ Mahalanobis Matrices that compute the distances between query and gallery feature vectors.
Each Mahalanobis Matrix is used for a specific combination of query and gallery illumination conditions.

\subsubsection{Illumination Switch}

We find the $N$ most common illumination labels of the target domain as described in Section \ref{subsec_2phase}.
For those target domain images that do not belong to these $N$ conditions, we need a way to assign their illumination labels.
The \textbf{Illumination Switch} in the SMB is used to classify the target domain images into one of these $N$ illumination conditions.
Similar to the Illumination Estimator in Section \ref{subsec_2phase}, we use ResNet-18 \cite{he_2016} as backbone to train a classifier network.
The synthetic images generated from different illumination conditions ($\mathbb{S}^{(1)}, \cdots, \mathbb{S}^{(N)}$) are used as training data, with images from $\mathbb{S}^{(n)}$ labeled as $n$ for $n \in \{1,2,\cdots,N\}$.

\subsubsection{Encoding Networks}\label{subsubsec_feature_extractor}

After estimating the illumination conditions using the \textbf{Illumination Switch}, the query image and the gallery image are assigned to the corresponding encoding networks for feature embedding.
For simplicity, we use ResNet-50 \cite{he_2016} as backbone of the encoding networks,
and each encoding network is trained as a classification problem as described in \cite{zheng_2016}. 
The synthetic images are used as training data and their identity labels are used as class labels.
Following the practice in \cite{sun_2017, zhong_2019}, we added two fully connected layers before the CrossEntropy layer to improve the training accuracy. 
The output of the first fully connected layer is 1024 dimensional and is followed by Batch-Normalization \cite{ioffe_2015}, ReLU, and Dropout \cite{srivastava_2014}.
The second fully connected layer is $1024 \times N_{\mathbb{S}}$, where $N_{\mathbb{S}}$ is the number of identities in the training set.
After the models are trained, we use the output of the first fully connected layer as the feature vectors.

We pre-train a baseline model using a combination of both real and virtual source domains (CUHK03 \cite{liwei_2014}, DukeMTMC4ReID \cite{ristani_2016}, and SyRI \cite{bak_2018}).
This combined domain consists of a very large number of identities (more than 3K) to represent the general identity features.
To adapt to a specific illumination condition in the target domain, each encoding network is fine tuned from this baseline model using the corresponding synthetic dataset.

\subsubsection{Mahalanobis Matrices}

Since feature vectors from different encoding networks may have different scales, we learn  Mahalanobis Matrices to compare feature vectors from different encoding networks.
To learn a Mahalanobis Matrix is to learn a global, linear transformation of the feature vectors \cite{koestinger_2012}, 
and the goal is to emphasize the relevant dimensions and diminish the irrelevant ones in the new feature space. 
For a set of feature vectors and their labels ($\{ (\mathbf{x_i}, y_i) \}_{i=1}^{K} $), a Mahalanobis Matrix, $\mathbf{M}$, computes the squared distance between two vectors $\mathbf{x_i}$ and $\mathbf{x_j}$ as:
\begin{equation}\label{eq_mdistance}
\begin{aligned}
d_{\mathbf{M}}^2(\mathbf{x_i}, \mathbf{x_j}) = (\mathbf{x_i} - \mathbf{x_j})^T \mathbf{M} (\mathbf{x_i} - \mathbf{x_j})
\end{aligned}
\end{equation}
\noindent and an algorithm to learn the Mahalanobis Matrix aims to minimize the distances within the same class while maximize the distance between different classes.
When $N=2$, we learn three Mahalanobis Matrices for different illumination conditions as shown in Figure \ref{fig_flowchart}: 
$\mathbf{M}^{(1)\times(1)}$ and $\mathbf{M}^{(2)\times(2)}$ are used when both query and gallery feature vectors have the same illumination condition, and $\mathbf{M}^{(1)\times(2)}$ is used when the illumination conditions are different.

There are several methods to find the Mahalanobis Matrix for a dataset through defining and optimizing an objective function \cite{davis_2007, guillaumin_2009, weinberger_2008, koestinger_2012}.
We choose the \textbf{K}eep \textbf{I}t \textbf{S}imple and \textbf{S}traightforward \textbf{ME}tric (KISSME) \cite{koestinger_2012} for its broad applicability on various benchmarks including person re-ID.
To learn the Mahalanobis Matrices, we use the feature vectors encoded from the synthetic images ($\mathbb{S}^{(1)}, \cdots, \mathbb{S}^{(N)}$) using the encoding networks.

\section{Experiments and Evaluation} \label{sec_evaluation}

\subsection{Re-ID Benchmark Datasets (Target Domain)}\label{subsec_dataset}

In the unsupervised re-ID scenario, we have access to one or more labeled domains (the ``source domains''), and unlabeled images from another domain that we wish to use. 
This new domain is known as the ``target domain''. 
In this paper, the experiments use 5 publicly available benchmark datasets as the target domains: PRID2011 \cite{hirzer_2011}, iLIDS-VID \cite{wang_2014}, Market-1501 \cite{zheng_2015}, CUHK01 \cite{li_2012}, and VIPeR \cite{gray_2007}. 
We follow the testing protocols in \cite{xiao_2016, paisitkriangkrai_2015} for splitting the target domains into query and gallery sets (details can be found below).
The re-ID performance of SMB for each target domain is evaluated and reported using rank-1 Cumulative Matching Characteristics (CMC-1) accuracy \cite{gray_2007}.\\
 
\noindent \textbf{PRID2011} 
contains images captured from two static surveillance cameras.
Camera A captured 385 identities, and Camera B captured 749 identities. 
We follow the single-shot settings and keep only one image for each identity from each camera. 
There are 200 identities that appeared in both cameras. 
For each experiment, half (100) of the 200 identities who appeared in both cameras are randomly chosen. 
Their images from Camera A are used as the query set, and their images from Camera B along with the images of the remaining 549 identities are used as the gallery set. 
There are 100/649 images in the query/gallery sets.\\

\noindent \textbf{VIPeR} 
contains 632 identities taken by two cameras with illumination and viewpoint variations. 
Each identity has one image from camera\_a and one image from camera\_b.
For each experiment, half (316) of the 632 identities are randomly chosen. 
The images from camera\_a are used as query set and the images from camera\_b as gallery set.
There are 316/316 images in the query/gallery sets.\\

\noindent \textbf{CUHK01} 
contains 971 identities in a campus environment that are captured by two surveillance cameras. 
There are two images for each identity in each camera, and a total of 3884 images in the dataset. 
For each experiment, 486 identities are randomly chosen. For each of the 486 identities, one random image from the first camera is used as the query image and one random image from the second camera as the gallery image. 
There are 486/486 images in the query/gallery sets.\\

\noindent \textbf{iLIDS-VID} 
contains 300 identities captured by two cameras. 
Each identity has one image from each camera (600 images in total). 
For each experiment, we randomly select 150 identities, and their images from camera 01/02 are used as the query/gallery set. 
There are 150/150 images in the query/gallery sets.\\

\noindent \textbf{Market-1501} contains 1501 identities and 32668 images from 6 cameras. 
The Market-1501 dataset is already split into two parts by the authors \cite{zheng_2015}: the first 751 identities and their 12936 images is used as the training set and the other 750 identities and their 19732 images as the testing set. 
The testing set is further split into the query and the gallery sets following the protocols in \cite{zheng_2015, yu_2017}. 
There are 3368/19732 images in the query/gallery sets.

\subsection{CycleGAN versus CycleDiffusion for Person Re-ID}\label{subsec_gan_diffusion}
We adopted two unpaired image-to-image translation methods for mapping images from the source domain to the target domain.
For CycleGAN \cite{zhu_2017}, we followed the network architecture from \cite{bak_2018}, and included the identity loss and masked reference loss in the training process.  
For CycleDiffusion, we used 3 resnet blocks in the U-Net architecture \cite{nichol_2021}. 
The diffusion step is set to be $T=1000$.
Each DDIM is trained for over 400k iterations.
The source domain $\mathbb{V}^{(n)}$ is the subset of the ULI-RI with the selected illumination label.
The target domain $\mathbb{T}$ is each real-life Re-ID benchmark dataset.
The synthetic images mapped from $\mathbb{V}^{(n)}$ to $\mathbb{T}$ is denoted as $\mathbb{S}^{(n)}$.

In Figure \ref{fig_prid_synthetic_samples}, we showed the example synthetic images generated using CycleGAN and CycleDiffusion. 
For both generative methods, the generated synthetic images have similar styles and illumination as the target domain, 
and the identity information are kept from the source domain such that they have the same identity labels as their source images.
Compared with CycleGAN, the synthetic images generated by CycleDiffusion have less distortion or missing part. 
We further used Frechet Inception Distance \cite{heusel_2017} to quantitatively evaluate the difference between the synthetic images and the target domain.
We used Peak Signal-to-Noise Ratio (PSNR) and Sturctural Similarity Index Measure (SSIM \cite{wang_2004}) to evaluate the similarity between the synthetic images and their corresponding source images.
For CycleDiffusion, we tested different values for the number of encoding steps. 
The quantitative evaluation results on the PRID dataset \cite{hirzer_2011} are reported in Table \ref{table_synthetic_evaluation}.

\begin{table}[t]
\centering
\caption{Quantitative Evaluation of CycleGAN and CycleDiffusion Using The PRID Dataset as The Target Domain. }\label{table_synthetic_evaluation}
\begin{adjustbox}{width=1\columnwidth}
\begin{tabular}{|c| c c c|c c c|}
\hline

 \multirow{2}{*}{ } &\multicolumn{3}{c|}{$\mathbb{V}^{(1)} \rightarrow \mathbb{T}$} &\multicolumn{3}{c|}{$\mathbb{V}^{(2)}\rightarrow \mathbb{T}$}\\  
 &FID$\downarrow$ &PSNR$\uparrow$ &SSIM$\uparrow$ &FID$\downarrow$ &PSNR$\uparrow$ &SSIM$\uparrow$ \\
\hline
 CycleGAN &77.65 &16.90 &0.343 &78.59 &17.82 &0.370\\
 \hline
\hline 
 $T_{es}=1000$   &68.45  &17.13 &0.401 &67.81 &16.97 &0.366\\
\hline
 $T_{es}=800$ &61.82  &19.65 &0.594 &\textbf{57.64} &18.90 &0.601\\
\hline
 $T_{es}=600$ &\textbf{59.03}  &19.66 &0.583 &58.74  &20.36 &0.577\\
\hline
 $T_{es}=400$ &62.44 &\textbf{21.33} &\textbf{0.732} &66.8 &\textbf{20.99} &\textbf{0.770}\\
\hline
\end{tabular}
\end{adjustbox}
\footnotesize{ \\The last four rows are the results of DDIM. $T_{es}$ is the number of encoding and decoding steps of CycleDiffusion. We do not include refine steps as described in \cite{wu_2023}. }
\end{table}

\subsection{Single-Illumination Model versus Synthesis Model Bank}

\begin{table}[t]
\centering
\caption{CMC-1 Accuracy of the Synthesis Model Bank and the Single-Illumination Models on Different Target Domains. The Synthetic Data for Training The Models Are Generated Using CycleGAN.}\label{table_single_model}
\begin{adjustbox}{width=0.9\columnwidth}
\begin{tabular}{|c|c|c|c|c|}
\hline
 Target Domain &Baseline &${\mathbb{S}}_{1}$ &${\mathbb{S}}_{2}$ &SMB\\
\hline
 PRID2011 &21.0 &43.0 &44.0 &\textbf{46.0}\\
\hline 
 VIPeR   &38.3  &46.1 &45.4 &\textbf{48.2}\\
\hline
 CUHK01 &50.8  &57.1 &56.1 &\textbf{57.8}\\
\hline
 iLIDS-VID &21.7  &36.2 &35.8 &\textbf{36.3}\\
\hline
 Market-1501 &58.4 &65.3 &64.3 &\textbf{66.8}\\
\hline
\end{tabular}
\end{adjustbox}
\footnotesize{ \\$\mathbb{S}_{1}$: The single-illumination model corresponds to the first illumination condition;
$\mathbb{S}_{2}$: The single-illumination model corresponds to the second illumination condition; 
SMB is the proposed Synthesis Model Bank.}
\end{table}

To demonstrate the effectiveness of using multiple illumination conditions, we compare the Synthesis Model Bank to the single-illumination model.
The single-illumination model consists of only one encoder and the corresponding Mahalanobis Matrix.
The Illumination Switch and the Mahalanobis Matrix for different illumination inputs (e.g., $\mathbf{M}^{(1)\times (2)}$ in Figure \ref{fig_flowchart}) are omitted.
The single-illumination model can be seen as a simplified version of the Synthesis Model Bank where we assume only one illumination condition in the target domain.

In the experiments, we use the entire target domain as test data for both the SMB and each single-illumination model.
For the single-illumination models, all test images are encoded by the same encoding network and compared by the the same Mahalanobis Matrix. 
The experiment results are reported using CMC-1 accuracy \cite{gray_2007}.
Table \ref{table_single_model} reports the results where the synthetic data are generated using CycleGAN.
We also include the results of the pre-trained baseline model as described in Section \ref{subsubsec_feature_extractor} for reference.
From Table \ref{table_single_model}, the two single-illumination models of each target domain both dramatically improve the re-ID accuracy compared with the baseline model.
The two single-illumination models also have similar performance on most target domains, possibly due to the fact that the numbers of images in the two illumination conditions are close.
Compared with the single-illumination models, the Synthesis Model Bank further improves the re-ID performance through integrating both illumination conditions.
Similar observations can be made when using CycleDiffusion to generate the synthetic data.

\subsection{Single-Illumination Model with the Ideal Single Illumination Input}

\begin{table}[t]
\begin{center}
\caption{Single-Illumination Model Tested with the Corresponding Single Illumination Input }\label{table_subdomain_model}
\begin{adjustbox}{width=1\columnwidth}
\begin{tabular}{|c|c|c|c|c|c|}
\hline
 Target   &Illumination &Query    &Gallery  &Valid   &CMC-1 \\
 Domain   &Condition    &Image    &Image    &Query   &Accuracy\\
\hline
\multirow{2}{*}{PRID2011}  &(1)  &51  &343 &27  &59.3 \\
\cline{2-6}
                           &(2)  &49  &306 &24  &54.2\\
\hline

\hline
\multirow{2}{*}{VIPeR}  &(1)  &149  &171 &97  &51.6\\
\cline{2-6}
                        &(2)  &167  &145 &93  &55.7\\
\hline

\hline
\multirow{2}{*}{CUHK01} &(1)  &400  &408 &345  &71.6\\
\cline{2-6}
                       &(2)  &86  &78 &23 &78.3 \\
\hline

\hline
\multirow{2}{*}{iLIDS-VID} &(1)  &100  &103 &67 &44.8\\
\cline{2-6}
                        &(2)  &50 &47 &15  &46.7\\
\hline

\hline
\multirow{2}{*}{Market-1501} &(1)  &2088  &11344 &2036 &76.1\\
\cline{2-6}
                       &(2)  &1280  &8388  &1218  &73.6\\
\hline
\end{tabular}
\end{adjustbox}
\end{center}
\end{table}

\begin{table*}[hbt!]
\centering
\caption{Comparison with Other Unsupervised Domain Adaption Methods for Person Re-ID }\label{table_sota}
\begin{adjustbox}{width=1.8\columnwidth}
\begin{tabular}{|c|c|c|c|c|c|c|c|}
\hline
  &Method &Source &PRID2011 &VIPeR &CUHK01 &iLIDS-VID &Market-1501\\
\hline
\hline

\multirow{4}{*}{\rotatebox[origin=c]{90}{$\quad$ FVO}}
&TL \cite{peng_2016} &Mixed\cite{peng_2016}  &24.2 &31.5 &27.1 &- &-\\
\cline{2-8}
&TJ-AIDL \cite{wangjingya_2018} &Market  &26.8 &38.5 &- &- &-\\
\cline{2-8}
&TJ-AIDL \cite{wangjingya_2018} &Duke &34.8 &35.1 &- &- &58.2\\
\cline{2-8}
&CAMEL \cite{yu_2017} &JSTL \cite{xiao_2016} &- &30.9 &57.3 &- &54.5\\
\hline

\hline
\multirow{4}{*}{\rotatebox[origin=c]{90}{SDA $\quad$}}
&PTGAN \cite{wei_2018} &Duke &- &- &- &- &38.6\\
\cline{2-8}
&PTGAN \cite{wei_2018}  &CUHK03 &37.5 &- &- &- &-\\
\cline{2-8}
&SyRI\cite{bak_2018} & $R^a+S^b$ &43.0 &43.0 &54.9 &- &65.7 \\
\cline{2-8}
&SPGAN+LMP \cite{deng_2018} &Duke &- &- &- &- &58.1 \\
\cline{2-8}
&CamStyle+LMP\cite{zhong_2019} &Duke &- &- &- &- &64.7 \\
\cline{2-8}
&SMB (CycleGAN) &$R^a+S^b$ &46.0 &48.2 &57.8 &36.3 &66.8 \\
\cline{2-8}
&\textbf{SMB (CycleDiffusion)} &$R^a+S^b$ &\textbf{47.7} &\textbf{49.0} &\textbf{60.6} &\textbf{37.1} &\textbf{69.9}\\
\hline

\end{tabular}
\end{adjustbox}\\
\rule{0pt}{2ex}  
\footnotesize{
$^a$ $R$ is the real source domain which includes CUHK03 \cite{liwei_2014} and DukeMTMC4ReID \cite{ristani_2016}. $^b$ $S$ is a virtual source domain SyRI \cite{bak_2018}. \\
``-'' means the target domain is not tested in the original paper.
All results are reported using the CMC-1 accuracy.
}
\end{table*}

To understand how the single-illumination models work within the Synthesis Model Bank, we evaluate each single-illumination model with test images that have the ``ideal'' illumination condition, i.e., the test images are chosen to have the same illumination condition as the training images of each encoding network. 
In these experiments, each target domain is split into 2 subsets according to the classification results of the Illumination Switch.
The target domain images which are classified as ``$l_{1}$'' (or ``$l_{2}$'') are assigned to the first (or the second) subset.
We call these two subsets $\mathbb{T}^{(1)}$ and $\mathbb{T}^{(2)}$ respectively with the superscripts indicating the illumination condition.
$\mathbb{T}^{(1)}$ is used to test the performance of the first single-illumination model, and analogously $\mathbb{T}^{(2)}$ for the second.
Note that in the original target domain, some query images and their matched gallery images have different illumination conditions,
and some query images in $\mathbb{T}^{(1)}$ and $\mathbb{T}^{(2)}$ do not have a matched gallery image in the same subset.
In these experiments, we exclude the query images that do not have a matched gallery image in the same subset.

Table \ref{table_subdomain_model} summarizes the experiment results.
Each target domain has two rows corresponding to the two single-illumination models (or illumination conditions).
The number of query and gallery images in each test case are reported in the third and the fourth column. 
The fifth column gives the number of valid query images, i.e., the query images that have a match in the gallery set.
Comparing the results in Table \ref{table_subdomain_model} to the results in Table \ref{table_single_model}, the single-illumination models with the ``ideal'' illumination input have better CMC-1 accuracy than the single-illumination models with the original target domain.
This shows that the single-illumination models have the best performance when the test images are in the same illumination condition.
The single-illumination models with the ``ideal'' illumination input also have better performance compared to the Synthesis Model Bank with the original target domain (last column of Table \ref{table_single_model}).
This gap of performance is likely to be caused by directly comparing the distances from different Mahalanobis Matrices in the Synthesis Model Bank.
For instance, a query image from $\mathbb{T}^{(1)}$ needs to be compared to the gallery images from both $\mathbb{T}^{(1)}$ and $\mathbb{T}^{(2)}$ in the Synthesis Model Bank. 
In this case, the distances from $\mathbf{M}^{(1)\times (1)}$ and the distances from $\mathbf{M}^{(1)\times (2)}$ are compared directly, which causes some correct retrievals within $\mathbf{M}^{(1)\times (1)}$ being misled by $\mathbf{M}^{(1)\times (2)}$.

\subsection{Comparison with Other Unsupervised Domain Adaption Methods for Person Re-ID}

Here we compare the proposed Synthsis Model Bank with other unsupervised domain adaption (UDA) methods for person re-identification.
To select related studies, we focus on unsupervised methods that carried out experiments on any of the target domains mentioned in Section \ref{subsec_dataset}.
These UDA methods for person re-ID can be categorized into three types: 
1) feature vector operation (FVO) \cite{wangjingya_2018, yu_2017, peng_2016} which uses CNN or hand-crafted methods to transfer feature vectors from different domains into a shared domain for comparing; 
2) synthesis data augmentation (SDA) \cite{bak_2018, wei_2018, deng_2018, zhong_2019} which generates synthetic images using GAN-based methods for fine tuning re-ID models; 
and 3) pseudo labeling (PL) \cite{bai_2021, ge_2020, zhao_2020, zou_2020, zhai_2020} which iteratively uses clustering methods to generate pseudo labels for the target domain images and re-train re-ID models. 

Table \ref{table_sota} summarizes the results of the reviewed methods. 
Among all FVO and SDA methods, the proposed Synthesis Model Bank achieves the best performance on most target domains. 
With the same source domains for pre-training the baseline model, our method outperforms the other SDA method by 4.2-6.0\% on various target domains.
On the specific target domain, Market-1501, most recent pseudo-labeling methods outperforms the synthesis-augmentation methods even with limited source domain (CMC-1 accuracy on Market-1501: UMSDA\cite{bai_2021}-$94.8\%$, MMT\cite{ge_2020}-$88.4\%$, NRMT\cite{zhao_2020}-$87.8\%$). 
Instead of indirectly learning information about the target domain through synthetic data, the pseudo-labeling methods directly use the target domain images during the iteratively training process. 
This indicates that there is still a gap between the distributions of the synthetic data and the target domain.
The pseudo-labeling methods are not included in Table \ref{table_sota} since most of them only used Market-1501 as target domain.
Using the small-scale re-ID benchmark datasets as target domain is not reported in the pseudo-labeling methods.

\section{Conclusion}
In this paper, we introduce the \textbf{U}E4 \textbf{L}abeled \textbf{I}mages for Person \textbf{R}e-\textbf{I}dentification (ULI-RI), a new 3D virtual-human dataset for person re-ID to better quantify the illumination intensity and improve the quality of CycleGAN synthesis for varied illumination intensities. 
This virtual human dataset provides rich diversity and has images labeled by illumination intensity as well as person identity, model z-rotation angle, and background. 
In addition, we propose a new deep learning architecture called the Synthesis Model Bank to address illumination variation in unsupervised person re-ID.
The Synthesis Model Bank makes use of synthetic data from different illumination conditions to train multiple CNNs and Mahalanobis matrices.
From our experiments, the Synthesis Model Bank can effectively improve the performance of single-illumination models and outperform other synthesis methods on several re-ID benchmarks. 
From the experiments of single-illumination models with ideal illumination input, it is possible to further improve the Synthesis Model Bank through jointly learned Mahalanobis matrices.

Still, the pseudo-labeling methods outperform synthesis augmentation methods on the specific target domain, Market-1501. 
This indicates the distribution of the synthetic data still differs from the target domain. 
Future work will include the pseudo-labeling methods on small-scale target domains.

\section*{Declaration of Interest}
The authors declare that they have no conflict of interest.

 {\small{} \bibliographystyle{ieee}
\bibliography{ref}
 }{\small\par}

\end{document}